\documentclass[prl,preprint,showpacs,preprintnumbers]{revtex4}
\usepackage{amsmath,amssymb,epsfig}

\begin{document}

\title{Flat band ferromagnetism without connectivity conditions in the 
flat band}
\author{Miklos~Gul\'acsi$^{}$, Gy\"orgy~Kov\'acs$^{}$, 
and Zsolt~Gul\'acsi$^{}$}
\address{
Department of Theoretical Physics, University of Debrecen, 
H-4010 Debrecen, Hungary}

\date{August 13, 2014}

\begin{abstract}
It is known that a system which exhibits a half filled lowest flat band and
the localized 
one-particle Wannier states on the flat band satisfy the connectivity 
conditions,
is always ferromagnetic. Without the connectivity conditions on
the flat band, the system is non-magnetic. We show 
that this is not always true. The reason is connected to a peculiar 
behavior of the band situated just above the flat band. 
\end{abstract}
\maketitle



Flat bands represent a real driving force nowadays since they 
appear in a broad class of subjects of large interest, as quantum Hall 
effect \cite{y0}, spin-quantum Hall effect \cite{y00}, 
topological phases \cite{y00,y1}, bose condensations \cite{y2}, highly
frustrated systems \cite{Intr22}, delocalization effects \cite{Intr30} or
symmetry broken ordered phases \cite{Intr15}. Among the ordered phases 
connected to flat bands, the flat band ferromagnetism \cite{Intr15,Intr15a} 
is the most important, providing a leading mechanism -- especially in 
organic or 
frustrated materials -- for the emergence of ferromagnetism in 
conditions in which magnetic atoms are completely missing from the system. 
In the mechanism of flat band ferromagnetism (on the lowest half filled bare
flat band), it is known that the system 
defined on a lattice (or graph) which can be described by a Hubbard type 
of model, 
is ferromagnetic for any arbitrary small on-site Coulomb repulsion $U > 0$,
if and only if the corresponding one-particle localized Wannier 
states are in contact with each other, i.e., the connectivity condition 
is satisfied for the bare flat band. 
If however, the connectivity condition for the localized one-particle states on
the flat band is not satisfied, hence the spins of the individual electrons
localized on the bare flat band are unable to correlate,  the system will
remain paramagnetic.

Several extensions of the original flat band ferromagnetism 
mechanism have been worked out. 
For example, ferromagnetism in the vicinity of flat bands
\cite{MT1}, or due to non-lowest energy bare flat bands
\cite{Intr24and26}, or on effective flat bands 
created by interaction in conditions in which bare flat bands are not present 
\cite{Intr25,Intr8,Intrx} and even in cases when large number of  
non-interacting sites are present in the system \cite{MT3}.

In this Letter we revisit the flat band ferromagnetism 
phenomenon and demonstrate rigorously, 
that even if the connectivity conditions are not satisfied for the
one-particle localized Wannier states on the bare lowest flat band, 
ferromagnetism is able to appear in the system. This is caused by an often
possible peculiar behavior of the dispersive band situated just above the
lowest flat band, which enforces the connectivity as will be detailed below.

\begin{figure}[b]
\includegraphics[width=6cm,height=4.25cm]{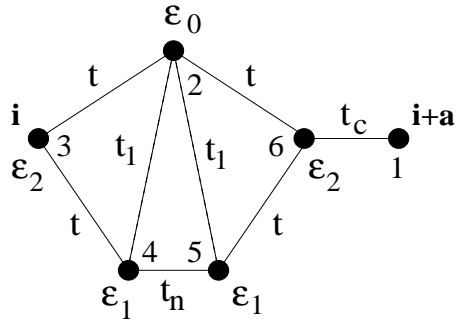}
\caption{The unit cell defined at the lattice site ${\bf i}$ of the 
pentagon chain under consideration. The numbers are representing the in-cell
notation of sites, $t,t_1,\epsilon_m$, m=0,1,2 are the hopping matrix elements
and the on-site one particle potentials present in $\hat H_0$, while
${\bf a}$ is the Bravais vector.}
\end{figure}

The technique we apply is based on positive semidefinite operator properties
which allowed us to work out exact results in systems and models 
where exact results were unheard of before, such as:
periodic Anderson model in one \cite{Orlik},
two \cite{Intr27}, or three \cite{Intr16and17}
dimensions; disordered and interacting systems in two dimensions
\cite{Intr28}; emergence of stripes and droplets in 2D \cite{Intr29}; 
delocalization effect caused by the on-site Coulomb interaction in 2D
\cite{Intr30}; exact results of non-integrable quadrilateral
\cite{Intr24and26} or pentagon \cite{Intr25,Intr8} chains.

The best way to explain this new effect is through the concrete example
of pentagon chains. We are studying pentagon chains because they are the 
building
blocks of a wide class of conducting polymers \cite{new01} and are a fascinating
class of materials with a wide range of applications \cite{new02-04}. These
pentagon chain polymers have been explored and utilized intensively in the
past \cite{new05-06}. In particular, polythiophene was studied in the search
for plastic ferromagnets and, more generally, for ferromagnetism in systems
made entirely of nonmagnetic elements. Just recently \cite{Intr25,MT3} we 
proved 
that ferromagnetism does exist in these class of polymers with the use of the
positive semidefinite operators. 

Hereafter, using the same technique, 
we study another class of pentagon chain polymers, namely
the poly(3-alkylthiophene) \cite{PAT}
which has not been studied before at all. The pentagon chain is
formed by blocks (see, Fig. 1) described by a Hubbard model containing on
each sites the on-site Coulomb repulsion $U > 0$. 
The non-interacting part of the Hamiltonian is: 
\begin{eqnarray}
\hat H_0 &=& \sum_{\sigma} \sum_{{\bf i}=1}^{N_c} \: \{ [  
t_c \hat c^{\dagger}_{{\bf i}+{\bf r}_6,\sigma} \hat c_{{\bf i}+{\bf a},
\sigma} + t_n \hat c^{\dagger}_{{\bf i}+{\bf r}_4,\sigma} \hat c_{{\bf i}+
{\bf r}_5,\sigma}
\nonumber \\
&+& t_1 (
\hat c^{\dagger}_{{\bf i}+{\bf r}_2,\sigma} \hat c_{{\bf i}+{\bf r}_4,\sigma} +
\hat c^{\dagger}_{{\bf i}+{\bf r}_2,\sigma} \hat c_{{\bf i}+{\bf r}_5,\sigma} )
+ t ( \hat c^{\dagger}_{{\bf i}+{\bf r}_2,\sigma} \hat c_{{\bf i},\sigma}
\nonumber\\
&+& 
\hat c^{\dagger}_{{\bf i}+{\bf r}_6,\sigma} \hat c_{{\bf i}+{\bf r}_2,\sigma}  +  
\hat c^{\dagger}_{{\bf i},\sigma}\hat c_{{\bf i}+{\bf r}_4,\sigma} +
\hat c^{\dagger}_{{\bf i}+{\bf r}_5,\sigma}\hat c_{{\bf i}+{\bf r}_6,\sigma}) 
+ H.c. ]
\nonumber\\
&+& \epsilon_0 \hat n_{{\bf i}+{\bf r}_2,\sigma} +
\epsilon_1 ( \hat n_{{\bf i}+{\bf r}_4,\sigma} + 
\hat n_{{\bf i}+{\bf r}_5,\sigma} ) 
\nonumber \\
&+&
\epsilon_2 (\hat n_{{\bf i},\sigma} + \hat n_{{\bf i}+{\bf r}_6,\sigma}) \} ,
\label{E1}
\end{eqnarray}
where $N_c$ represents the number of cells. The sites inside the unit cell
constructed at the lattice site ${\bf i}$ are placed at ${\bf i}+{\bf r}_n$,
where $n=2,3,...6$ represents the in-cell notation of sites, the $n=1$ value
denotes the ${\bf i}+{\bf a}$ site where ${\bf a}$ is the Bravais vector,
and for mathematical convenience ${\bf r}_3=0$ is considered.
The one-particle on-site potentials are denoted by $\epsilon_0$ on the site 
${\bf i}+{\bf r}_2$; $\epsilon_1$ on sites ${\bf i}+{\bf r}_4$ 
${\bf i}+{\bf r}_5$; while on the sites
${\bf i}$ and ${\bf i}+{\bf r}_6$ (hence also on ${\bf i}+{\bf a}$ 
in the next cell) by $\epsilon_2$. The nearest neighbor
hopping matrix elements are  
$t_{4,5}=t_n$ on the lower horizontal bond of the cell; $t_{6,1}=t_c$ on the 
horizontal external connecting bond of the cell, and
$t_{3,4}=t_{5,6}=t_{2,3}=t_{6,2}=t$ on the circumference of the pentagon.
In the poly(3-alkylthiophene) polymers there is 
also a next-nearest neighbor hopping, denoted by $t_1$, because at the
site $n=2$ there is always a bigger atom present
relative to sites $n=3,4,5,6$. Note that
we have $m=5$ sites per unit cell, hence 5 sub-lattices are present providing 5
bands in the system. 

The full Hamiltonian is $\hat H=\hat H_0 + \hat H_U$, where the interacting
part is given by $\hat H_U=\sum_{{\bf i}=1}^{N_c}\sum_{n=2}^6 U_{{\bf i}+
{\bf r}_n}\hat n_{{\bf i}+
{\bf r}_n,\uparrow} \hat n_{{\bf i}+{\bf r}_n,\downarrow}$, $U_{\bf j}=U > 0$ for all
${\bf j}$.

In order to find the band structure of $\hat H_0$ we Fourier
transform the Fermi operators from the Hamiltonian via
$\hat c_{{\bf i}+{\bf r}_n,\sigma}= (1/\sqrt{N_c}) \sum_{{\bf k}=1}^{N_c}
e^{-i{\bf k}{\bf i}} e^{-i {\bf k}{\bf r}_n} \hat c_{n,{\bf k},\sigma}$,
where ${\bf k}$ is directed along the line of the chain, and one has
$|{\bf k}|= k = 2 m \pi/(a N_c)$, $m=0,1,2,...,N_c-1$, $|{\bf a}|=a$ being the
lattice constant. After this step, the non-interacting part of the Hamiltonian 
becomes
\begin{eqnarray}
\hat H_0= \sum_{\sigma} \sum_{k=1}^{N_c} (\hat c^{\dagger}_{2,k,\sigma},
\hat c^{\dagger}_{3,k,\sigma}, ..., \hat c^{\dagger}_{6,k,\sigma} ) \tilde M
\left( \begin{array}{c}
\hat c_{2,k,\sigma} \\
\hat c_{3,k,\sigma} \\
.....               \\
\hat c_{6,k,\sigma} \\
\end{array} \right) ,
\label{E2}
\end{eqnarray}
where the $5\times 5$ matrix $\tilde M$ is:
\begin{widetext}
\begin{eqnarray}
\tilde M =
\left( \begin{array}{ccccc}
\epsilon_0 & t e^{+ i \frac{kb}{2}} & t_1e^{ik(\frac{b}{2}-b_2)} & t_1e^{-ik(\frac{b}{2}-b_2)} 
& t e^{-i \frac{k b}{2}} \\
t e^{-i \frac{k b}{2} } & \epsilon_2 & t e^{-i k b_2} & 0 & t_c e^{i k b'} \\
t_1e^{-ik(\frac{b}{2}-b_2)} & t e^{+ i k b_2} & \epsilon_1 & t_n e^{- i k b_1} & 0 \\
t_1e^{ik(\frac{b}{2}-b_2)} & 0 & t_n e^{+ i k b_1} & \epsilon_1 & t e^{- i k b_2} \\
t e^{+i \frac{k b}{2}} & t_c e^{- i k b'} & 0 & t" e^{+ i k b_2} & \epsilon_2
\end{array} \right) .
\nonumber
\end{eqnarray}
\end{widetext}
Here distances $b_{\alpha}$ are expressed by 
the unit vector ${\bf u}$ directed along ${\bf k}$, obtaining
$b_1=|{\bf r}_5-{\bf r}_4|, b_2=|({\bf r}_4-{\bf r}_3){\bf u}|, 
b'=|{\bf a}-{\bf r}_6|, b=|{\bf r}_6-{\bf r}_3|$, 
$a=b+b'$, and $b=b_1+2b_2$. The band structure in obtained by 
diagonalizing $\tilde M$. 

This yields $E_{\bf k}=\epsilon$ from
the equation $0 = A + B \cos (ak)$, where
\begin{eqnarray}
A&=& [ (\epsilon_0-\epsilon)(\epsilon_2-\epsilon)-2 t^2 ]
\nonumber \\
&\times&[ (\epsilon_1-\epsilon)^2(\epsilon_2-\epsilon) - t^2(\epsilon_1-\epsilon)
-t_n^2(\epsilon_2-\epsilon)] 
\nonumber\\
&+& 2 [t^2 -t_1(\epsilon_2-\epsilon)]
\nonumber \\
&\times& [(\epsilon_1-\epsilon)(\epsilon_2-\epsilon)t_1 +t_n t^2
- t_1 t^2- (\epsilon_2-\epsilon) t_1 t_n ]
\nonumber\\
&+&t \: t_1 \{ t \big[(\epsilon_1-\epsilon)(\epsilon_2-\epsilon)-t^2\big]
\nonumber \\
&+& t [(\epsilon_1-\epsilon)(\epsilon_2-\epsilon)-t^2\big] \}
\nonumber \\
&-& t^2 (\epsilon_0-\epsilon) \big[(\epsilon_1-\epsilon) (\epsilon_2-\epsilon)-
t^2 ]
\nonumber\\
&+& 2t_1^2 t_c^2 \big[(\epsilon_1-\epsilon) -t_n \big]
-(\epsilon_0-\epsilon)t_c^2 \big[(\epsilon_1-\epsilon)^2 -t^2_n \big], 
\nonumber\\
\end{eqnarray}
and 
\begin{eqnarray}
B&=&2 \{ 
t^2 t_c [ (\epsilon_1 - \epsilon)^2-t_n^2 ]
+ 2t \: t_c t_1 [ t_n t - t (\epsilon_1 - \epsilon)]
\nonumber \\
&-& t^2 t_c \big[t_n(\epsilon_0 -\epsilon)-t_1^2 \big] \}.
\label{E4}
\end{eqnarray}

From this the flat band condition is obtained when simultaneously $A=0$ and $B=0$:
\begin{eqnarray}
\epsilon_0 &=& 2(\epsilon_1-t_n)+ \frac{(t_1-\epsilon_1+t_n)^2}{t_n}, 
\nonumber \\
\epsilon_2 &=& \frac{t^2}{\epsilon_1-t_n}+\frac{t_c^2(\epsilon_1-t_n)}{
\epsilon_2(\epsilon_1-t_n)-t^2}.
\label{E5}
\end{eqnarray}
In order to place the flat band in lowest position, supplementary conditions
must be imposed, which can written as
\begin{eqnarray}
&& \epsilon_0, \epsilon_1, \epsilon_2 > 0, \quad t_n > 0, 
\nonumber \\
&& \epsilon_1 -t_n > 0, \quad
\epsilon_2(\epsilon_1-t_n)-t^2 > 0.
\label{E6}
\end{eqnarray}

The ground state on the lowest flat band can be easily constructed by 
transforming the starting Hamiltonian in positive semidefinite form. In the
case of $m=5$ sub-lattices, the use of $\: m-1=4$ block operators for this 
transformation always lead to transformation conditions which provide a 
flat band \cite{Intrx}. In the present case the used four block operators as
linear combinations of fermionic operators acting on the sites of the block,
are defined for each unit cell on three triangles and a bond as follows
\begin{eqnarray}
&&\hat A_{1,{\bf i},\sigma}= a_{1,2} \hat c_{{\bf i}+{\bf r}_2,\sigma} +
a_{1,3} \hat c_{{\bf i}+{\bf r}_3,\sigma} +
a_{1,4} \hat c_{{\bf i}+{\bf r}_4,\sigma} , 
\nonumber \\
&&\hat A_{2,{\bf i},\sigma}= a_{2,2} \hat c_{{\bf i}+{\bf r}_2,\sigma} +
a_{2,4} \hat c_{{\bf i}+{\bf r}_4,\sigma} +
a_{2,5} \hat c_{{\bf i}+{\bf r}_5,\sigma} ,
\nonumber\\  
&&\hat A_{3,{\bf i},\sigma}= a_{3,2} \hat c_{{\bf i}+{\bf r}_2,\sigma} +
a_{3,5} \hat c_{{\bf i}+{\bf r}_5,\sigma} +
a_{3,6} \hat c_{{\bf i}+{\bf r}_6,\sigma} , 
\nonumber \\
&&\hat A_{4,{\bf i},\sigma}= a_{4,6} \hat c_{{\bf i}+{\bf r}_6,\sigma} +
a_{4,1} \hat c_{{\bf i}+{\bf a},\sigma} ,
\label{E7}
\end{eqnarray}
where the coefficients $a_{i,j}$ denote the numerical prefactor of the
Fermi operator from the block operator $i$ at the site ${\bf r}_j$.
The transformation in positive semidefinite form leads to
\begin{eqnarray}
\hat H= \hat H_A + \hat H_U, \quad
\hat H_A =\sum_{\sigma} \sum_{{\bf i}=1}^{N_c} \sum_{\alpha=1}^4
\hat A^{\dagger}_{\alpha,{\bf i},\sigma} \hat A_{\alpha,{\bf i},\sigma}. 
\label{E8}
\end{eqnarray}
The matching equations are (note that periodic boundary conditions are used):
\begin{eqnarray}
&&t_n=a^*_{2,4} a_{2,5}, \quad t_c=a^*_{4,6} a_{4,1}, 
\nonumber \\
&&t = a^*_{1,2} a_{1,3} = a^*_{3,6} a_{3,2}=a^*_{1,3} a_{1,4}=  a^*_{3,5} a_{3,6},
\nonumber\\
&&t_1= a^*_{2,2} a_{2,5} + a^*_{3,2} a_{3,5}= a^*_{2,2} a_{2,4} + a^*_{1,2} a_{1,4} ,
\nonumber \\
&&\epsilon_0 = |a_{1,2}|^2 + |a_{3,2}|^2 +|a_{2,2}|^2,
\nonumber\\
&&\epsilon_1= |a_{1,4}|^2 + |a_{2,4}|^2=  |a_{2,5}|^2 + |a_{3,5}|^2, 
\nonumber \\
&&\epsilon_2= |a_{1,3}|^2 + |a_{4,1}|^2 =  |a_{3,6}|^2 + |a_{4,6}|^2 .
\label{E9}
\end{eqnarray}
The equations (\ref{E9}) lead to the solution
\begin{eqnarray}
&&a_{1,2}=a_{1,4}=a_{3,2}=a_{3,5}=sign(t) \sqrt{\epsilon_1-t_n}e^{i\phi_1}, 
\nonumber \\
&&a_{1,3}=a_{3,6}=\frac{|t|}{\sqrt{\epsilon_1-t_n}} e^{i\phi_1},
\nonumber\\
&&a_{2,4}=a_{2,5}=\sqrt{t_n}e^{i\phi_2}, \quad 
a_{2,2}=\frac{t_1-\epsilon_1+t_n}{\sqrt{t_n}}e^{i\phi_2},
\nonumber\\
&&a_{4,1}=\sqrt{\frac{\epsilon_2(\epsilon_1-t_n)-t^2}{\epsilon_1-t_n}}
e^{i\phi_3}, 
\nonumber \\
&&a_{4,6}=t_c \sqrt{ \frac{\epsilon_1-t_n}{\epsilon_2
(\epsilon_1-t_n)-t^2}}e^{i\phi_3},
\label{E10}
\end{eqnarray}
where $\phi_{\alpha}$, $\alpha=1,2,3$ are arbitrary phases.
The conditions under which (\ref{E9}) has solutions, and that the obtained
flat band is in the lowest position coincide to
the conditions in (\ref{E5},\ref{E6}).

Once the block operators from (\ref{E7}) are worked out, the ground state can be
easily constructed from the new block operators $\hat B^{\dagger}_{\alpha_{\bf i},
\sigma_{\bf i}}$
defined on all lattice sites ${\bf i}$ as
\begin{eqnarray}
\hat B^{\dagger}_{\alpha_{\bf i},\sigma_{\bf i}}=\sum_{{\bf j}=1}^{N_c}\sum_{n=2}^6 
x_{{\bf j},n} \hat c^{\dagger}_{{\bf j}+{\bf r}_n,\sigma_{\bf i}}.
\label{E11}
\end{eqnarray}
Here, $\alpha_{\bf i}$ is an index denoting linearly independent $\hat B^{\dagger}$
operators. The requirement for the $\hat B^{\dagger}_{\alpha_{\bf i},\sigma_{\bf i}}$ 
operators
is: i) to satisfy for all values of all indices the anti-commutation relations
\begin{eqnarray}
\{ \hat A_{n,{\bf i}',\sigma'}, \hat B^{\dagger}_{\alpha_{\bf i},\sigma_{\bf i}} \}
=0,
\label{E12}
\end{eqnarray}
and ii) the product $\prod_{\bf i}\hat B^{\dagger}_{\alpha_{\bf i},\sigma_{\bf i}} 
|0\rangle$, where $|0\rangle$ is the bare vacuum, must not introduce double 
occupancy in the system. In this case, at half filling lowest flat band
the ground state becomes
\begin{eqnarray}
|\Psi_g \rangle = \prod_{{\bf i}=1}^{N_c} \hat B^{\dagger}_{
\alpha_{\bf i},\sigma_{\bf i}} |0\rangle .
\label{E13}
\end{eqnarray}
This indeed satisfies property i) since the relation 
$\hat H_A |\Psi_g\rangle =0$
is satisfied, and due to the property ii), also $\hat H_U |\Psi_g\rangle =0$
holds. The uniqueness proof can be easily done, for example, on the line of 
\cite{Intr8}.
 
On the bare flat band the one-particle states are
given by $|\phi_{{\bf i},\sigma_{\bf i}}\rangle=\hat B^{\dagger}_{
\alpha_{\bf i},\sigma_{\bf i}} |0\rangle$. Usually these are localized states
because the solutions of (\ref{E12}) provide $\hat B^{\dagger}_{\alpha_{\bf i},
\sigma_{\bf i}}$ operators which act only on the sites of a finite block.

The connectivity conditions exist if the neighboring $\hat B^{\dagger}_{
\alpha_{\bf i},\sigma_{\bf i}}$ operators are in contact with each other 
at least on one site. If this property however is missing, the localized 
one-particle states on the flat band are called ``disconnected'', i.e., do not 
satisfying the connectivity condition.

There is a great difference between the physical properties of the ground state
(\ref{E13}) with connectivity, or without connectivity conditions. This is
because when connectivity exists, the  $\hat B^{\dagger}_{
\alpha_{\bf i},\sigma_{\bf i}}$ operators are in contact with each other. 
Thus, in order to not have
double occupancy (condition ii) under (\ref{E11})), i.e., to avoid
the increase in energy caused by the Hubbard interaction, the system must
fix the spin index of all operators in (\ref{E13}). Consequently the system
becomes ferromagnetic. This is the flat band ferromagnetism phase. 
Contrary to this, when the
connectivity condition does not exist and the different $\hat B^{\dagger}_{
\alpha_{\bf i},\sigma_{\bf i}}$ operators are not in contact with each other, 
the spin on individual $\hat B^{\dagger}_{\alpha_{\bf i},\sigma_{\bf i}}$ 
operators can remain arbitrary. This is due to the fact that, double occupancy 
does not occur and hence the Hubbard interaction is completely avoided. In 
this case the $|\Psi_g\rangle$ ground state from (\ref{E13}) becomes paramagnetic.

\begin{figure}[b]
\includegraphics[width=6cm,height=4.6cm]{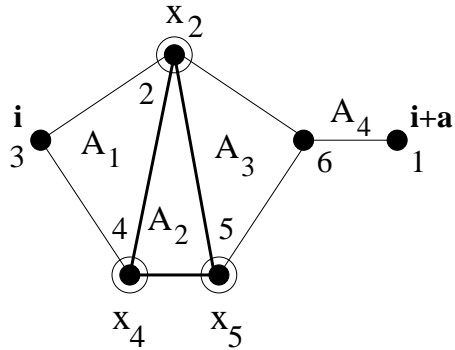}
\caption{The block on which the block operator $\hat B^{\dagger}_{
\alpha_{\bf i},\sigma_{\bf i}}$ is defined at the  
lattice site ${\bf i}$ when connectivity not exists. 
The block is presented with thick lines, has the
form of a triangle, and it contains (see circles) the sites (2,4,5).
The coefficients $x_2,x_4,x_5$ are the prefactors [see (\ref{E11})]
of the sites present in the block.}
\label{Fig2}
\end{figure}

In the present case non-connected $\hat B^{\dagger}_{\alpha_{\bf i},\sigma_{\bf i}}$
block operators can appear as defined on internal triangle blocks in each 
cell, as depicted in Fig.2. If such a type of solution of (\ref{E12}) exists,
it must exist for all ${\bf j}$ lattice sites
\begin{eqnarray}
&&x_2=x_{{\bf j},2} \ne 0, \quad
x_4=x_{{\bf j},4} \ne 0, 
\nonumber \\
&&x_5=x_{{\bf j},5} \ne 0, \: \:  
x_{{\bf j},n =3,6}=0.
\label{E14}
\end{eqnarray}
In this case the  $\hat B^{\dagger}_{\alpha_{\bf i},\sigma_{\bf i}}$ operators become
\begin{eqnarray}
\hat B^{\dagger}_{{\bf i},\sigma_{\bf i}} &=&
\hat B^{\dagger}_{\alpha_{\bf i},\sigma_{\bf i}}
\nonumber \\
&=& x_2 \hat c^{\dagger}_{{\bf i}+{\bf r}_2,\sigma_{\bf i}}+
x_4 \hat c^{\dagger}_{{\bf i}+{\bf r}_4,\sigma_{\bf i}}+
x_2 \hat c^{\dagger}_{{\bf i}+{\bf r}_2,\sigma_{\bf i}}.
\label{E15}
\end{eqnarray}
In order to satisfy (\ref{E14},\ref{E15}), the equation (\ref{E12}) gives
the system of equations
\begin{eqnarray}
&&a_{1,2} x_2 + a_{1,4} x_4 = 0,
\nonumber\\
&&a_{3,2} x_2 + a_{3,5} x_5 = 0,
\nonumber\\
&&a_{2,2} x_2 + a_{2,4} x_4 + a_{2,5} x_5 = 0 ,
\label{E16}
\end{eqnarray}
which provides $x_2,x_4,x_5 \ne 0$ nontrivial solution only if
\begin{eqnarray}
t_1=\epsilon_1+|t_n|.
\label{E17}
\end{eqnarray}
The relation (\ref{E17}) becomes a supplementary condition leading to the
non-connectivity of the localized one-particle states on the flat band. It can 
be easily checked that when (\ref{E17}) is satisfied, and $x_2,x_4,x_5 \ne 0$
holds, we automatically have
$x_{{\bf j},3}=x_{{\bf j},6}=0$, consequently, connected solutions do not exist.

We further note, that when the condition (\ref{E17}) is satisfied (and given
by (\ref{E6}), the inequality $t_n > 0$ holds), Eq.(\ref{E5}) which represents
the flat band condition, becomes
\begin{eqnarray}
\epsilon_0=2(\epsilon_1-t_n)+4t_n, \quad \epsilon_2=\frac{t^2}{\epsilon_1-t_n}
+ |t_c|.
\label{E18}
\end{eqnarray}
In summary, when (\ref{E6},\ref{E17},\ref{E18})
are all satisfied, the lowest flat band with one-particle localized
states that extends over triangles that do not connect with each other in
each unit cell (see Fig.2), 
the connectivity condition is not satisfied. In this conditions, 
at half filling lowest flat band,
the ground state must be of the form (see (\ref{E13},\ref{E15}))
\begin{eqnarray}
|\Psi_g \rangle = \prod_{{\bf i}=1}^{N=N_c} \hat B^{\dagger}_{{\bf i},\sigma_{\bf i}} 
|0\rangle,
\label{E19}
\end{eqnarray}
Here, $\sigma_{\bf i}$ in each cell is arbitrary, i.e.
the ground state is non-magnetic \cite{Explic1}.

However, the physical ground state is not of the form (\ref{E19}), 
i.e., the wave vector (\ref{E19}) does not span the kernel of (\ref{E8}),
consequently the uniqueness of (\ref{E19}) as the ground state of the
Hamiltonian given in (\ref{E8}) cannot be demonstrated. 

The reason why the ground state from (\ref{E19}) is not the true 
ground sate is as follows.
The number $N_c$ of linearly independent operators  $\hat B^{\dagger}_{{\bf i},
\sigma_{\bf i}}$ which were deduced from (\ref{E12}) when the non-connectivity
condition (\ref{E17}) holds, are not forming the complete set of
solutions of (\ref{E12}). That is, there
exists another linearly independent $\hat B^{\dagger}_{\sigma}$ operator satisfying
(\ref{E12}) when (\ref{E17}) together with (\ref{E6}) and (\ref{E18}) holds. 
This operator is extended, and is not related to the states in the
lowest flat band. At $t_c>0$ it has the form
\begin{eqnarray}
\hat B^{\dagger}_{1,\sigma} =  \sum_{\bf i} 
[a(\hat c^{\dagger}_{{\bf i}+{\bf r}_4,\sigma} 
&-& \hat c^{\dagger}_{{\bf i}+{\bf r}_5,\sigma}) 
\nonumber \\
+ b(\hat c^{\dagger}_{{\bf i}+{\bf r}_6,\sigma}
&-& \hat c^{\dagger}_{{\bf i}+{\bf r}_1,\sigma})],
\label{E20}
\end{eqnarray}
while at $t_c < 0$ it can be expressed as
\begin{eqnarray}
\hat B^{\dagger}_{2,\sigma}=  \sum_{\bf i} (-1)^i 
[a(\hat c^{\dagger}_{{\bf i}+{\bf r}_4,\sigma}
&-& \hat c^{\dagger}_{{\bf i}+{\bf r}_5,\sigma}) 
\nonumber \\
+ b(\hat c^{\dagger}_{{\bf i}+{\bf r}_6,\sigma}
&+& \hat c^{\dagger}_{{\bf i}+{\bf r}_1,\sigma})],
\label{E21}
\end{eqnarray}
where in the last relation $i=|{\bf i}|/|{\bf a}|$ is an integer number
which represents the length of the vector ${\bf i}$ in lattice constant units.
In both cases, $b/a=a_{3,5}/a_{3,6}=a_{1,4}/a_{1,3}$ holds for the numerical
prefactors in (\ref{E20},\ref{E21}). We point out that
in the process of  computing these results, when (\ref{E6},\ref{E17},\ref{E18})
holds, $a_{4,1}=a_{4,6}/sign(t_c)=\sqrt{|t_c|}e^{i\phi_3}$ is obtained.
For example, a sketch
of the $\hat B^{\dagger}_{1,\sigma}$ operator is shown in Fig.3.

\begin{figure}[b]
\includegraphics[width=8cm,height=1.41cm]{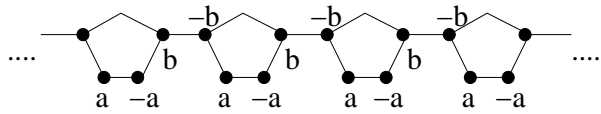}
\caption{The infinite block providing the extended operator
$\hat B^{\dagger}_{1,\sigma}$ being present in all cells 
at $t_c > 0$. The black dots denote the sites present in the 
block, while the coefficients represent the numerical prefactors of the
creation operators acting on the given site.}
\label{Fig3}
\end{figure}

The study of the eigenvectors of $\tilde M$ shows that the one-particle
extended states $|\phi_{\gamma}\rangle = \hat B^{\dagger}_{\gamma,\sigma} |0\rangle$,
$\gamma=1,2$ are the $ka=0$ (for $\gamma=1$, note that in this case $t_c>0$), 
and $ka=\pi$ (for $\gamma=2$, case in which $t_c<0$) eigenstates of the 
dispersive band situated just above the flat band. Since these can be obtained 
from (\ref{E12}), it points to the fact that these states have energy equal to the flat
band. In other terms, for $t_c>0$, the dispersive band situated
just above the flat band is in contact with the flat band at ${\bf k}=0$, while for
$t_c<0$ at ${\bf k}{\bf a}=\pi$.  A sketch 
at $t_c > 0$ is shown in Fig.4.

\begin{figure}[b]
\includegraphics[width=5cm,height=6.66cm]{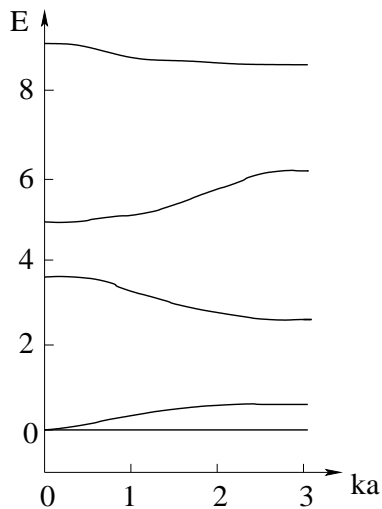}
\caption{The band structure of $\hat H_0$ taken at $t=1$, $t_n=1.2$,
$t_c=1.3$, $\epsilon_1=1.5$, when the conditions (\ref{E6},\ref{E17},\ref{E18})
are satisfied. Since $t_c > 0$, the dispersive band placed just above the
flat band is in contact with the flat band at ${\bf k}=0$. Note that the energy is an
even function of $ka$, hence only the $ka > 0$ part of the band structure 
is plotted.}
\label{Fig4}
\end{figure}

Following these considerations, the physical ground state becomes as follows.
We use, for example, the $t_c > 0$ case (at $t_c <0$ the operator
$\hat B^{\dagger}_{1,\sigma}$ has to be replaced in all equations by 
$\hat B^{\dagger}_{2,\sigma}$). In these conditions, 
at $N=N_c+1$ number of electrons we obtain
\begin{eqnarray}
|\Psi_g (N_c+1)\rangle = \hat B^{\dagger}_{1,\sigma} \prod_{{\bf i}=1}^{N_c} 
\hat B^{\dagger}_{{\bf i},\sigma} |0\rangle.
\label{E22}
\end{eqnarray}
Note that the $\sigma$ index is fixed. This is because 
$\hat B^{\dagger}_{1,\sigma}$,
given by the upper dispersive band, enforces the connectivity by being in 
contact with 
all operators $\hat B^{\dagger}_{{\bf i},\sigma}$. Consequently, the ground state is
ferromagnetic, even if the one-particle localized states on the flat band does
not satisfy the connectivity condition. Note that at $N_c >> 1$,
the experimental concentration connected to (\ref{E22}) in fact
corresponds to the half
filled lowest band. 

At $N=N_c$ number of electrons we obtain the following structure for 
$|\Psi_g\rangle$:
\begin{eqnarray}
|\Psi_g (N_c)\rangle &=& \sum_{i=1}^{N_c+1} b_i \: [ \hat B^{\dagger}(1,\sigma_1)
B^{\dagger}(2,\sigma_2) \ldots
\nonumber \\
&\times& \hat B^{\dagger}(i-1,\sigma_{i-1})
B^{\dagger}(i+1,\sigma_{i+1}) \ldots
\nonumber \\
& &
\nonumber \\
&\times& \hat B^{\dagger}(N_c,\sigma_{N_c})
\hat B^{\dagger}(N_c+1,\sigma_{N_c+1})] |0\rangle,
\label{E23}
\end{eqnarray}
where we have denoted in order, the operators $\hat B^{\dagger}_{1,\sigma},
\hat B^{\dagger}_{{\bf i}_1,\sigma},\hat B^{\dagger}_{{\bf i}_2,\sigma}, ...
\hat B^{\dagger}_{{\bf i}_{N_c},\sigma}$, by the operators present in the set 
${\cal{S}}=[\hat B^{\dagger}(1,\sigma),\hat B^{\dagger}(2,\sigma),...,
\hat B^{\dagger}(N_c+1,\sigma)]$. Note that 
the sum in (\ref{E23}) contains $N_c+1$ terms. All terms contain a product of
$N_c$ operators taken from the set ${\cal{S}}$, such that an arbitrary operator
with index $i$ from ${\cal{S}}$ is missing. The numerical prefactor $b_i$
holds the index of the missing operator. Note that only the first term from
(\ref{E23}) does not satisfy the connectivity condition (hence it has
a product of $N_c$ operators with arbitrary spin projection). For all other
$N_c$ terms $i > 1$ in (\ref{E23}), containing each a product of $N_c$ operators
taken from ${\cal{S}}$, one has $\sigma_1=\sigma_2=...=\sigma_{N_c+1}=\sigma$,
so the spin projection is fixed. This is enforced by the connectivity condition
effective in the $i>1$ terms, introduced by the $\hat B^{\dagger}_{1,\sigma}$ 
operator present in all these contributions. Since only one term from $N_c+1$
in (\ref{E23}) has arbitrary spin projections, the ground state 
$|\Psi_g(N_c)\rangle$ represents also a ferromagnetic state at $N_c >> 1$
\cite{obs1}, 
where $N=N_c$ corresponds exactly to the half filled lowest band. (Note
that, there are no physical reasons why huge differences should
be in a realistic cases between the magnitudes of different $|b_i|$
contributions.)

At the electron number $N < N_c$ the ground state becomes of the form
\begin{eqnarray}
|\Psi_g(N < N_c)\rangle &=& \sum_{\cal{D}} \alpha_{(i_1,i_2,\ldots,i_N)}
[\hat B^{\dagger}(i_1,\sigma_{i_1})
\nonumber \\
&\times& \hat B^{\dagger}(i_2,\sigma_{i_2}) \ldots
\hat B^{\dagger}(i_N,\sigma_{i_N})] |0\rangle,
\label{E24}
\end{eqnarray}
where ${\cal{D}}=\{ i_1,i_2,...,i_N \}$ is the set of all possible combinations
of the integers $(i_1,i_2,...,i_N)$ labeling the components of ${\cal{S}}$.
In this expression the sum has $p_1=C^{N}_{N_c+1}$ terms \cite{Explic2}, and from
these only $p_2=C^{N-1}_{N_c}$ contributions have connectivity conditions, where
$p_2/p_1=N/(N_c+1)$. Hence, with decreasing $N<N_c$ and increasing $N_c$ the 
ferromagnetism disappears below the half filled lowest band. The analysis of 
this 
crossover exceeds the frame of the present Letter and will be discussed 
elsewhere.

We mention that the ground states from (\ref{E22},\ref{E23},\ref{E24}) being
constructed at the mentioned $N$ with the complete set of solutions of 
(\ref{E12}) are unique since span the kernel of (\ref{E8}). The results remain
valid even if only sites ${\bf i}+{\bf r}_4$ or ${\bf i}+{\bf r}_5$ are only
interacting (hence 80 \% of sites are non-interacting in the system). We 
further note that stability studies made for flat band ferromagnetism before 
not include the here presented case \cite{obs2}.

In conclusions, in 
a system in which there is a lowest bare flat band of one-particle
localized states which do not satisfy the connectivity condition, 
the flat band ferromegnetism does not work and the ground state 
of the half filled lowest band is not a ferromagnet. 
Contrary to this, we rigorously proved that, in some circumstances
ferromagnetism is still possible. The reason for this is that
the dispersive band which appears just above the lowest flat band 
can be forced to be in contact with 
the lowest flat band. This contact point represents a 
particular extended one particle state which belongs to the dispersive band, 
but 
which has the energy of the one particle states from the flat band. This state
being extended, will introduce the connectivity condition, enforcing a
ferromagnetic state. We showed that this phenomenon exists in a class of
pentagon chains in which the conditions leading to the lowest
flat band containing the non-connected localized one-particle states, 
automatically
leads to one contact point with the dispersive band situated just above the
flat band. 

We underline that this is not a rare effect. Indeed, for a kinetic
Hamiltonian $\hat H_0$ with several hopping matrix elements $t_{\nu}$ and one 
particle potentials $\epsilon_{\nu}$, the bare band energies $\epsilon$ can be 
obtained usually from a relation of the type 
$A(\{\epsilon,t_{\nu},\epsilon_{\nu}\})+ B(\{\epsilon,t_{\nu},\epsilon_{\nu}\})  
\cos{\bf k}{\bf a}=0$ \cite{Explic3}, and the minimum
distance (i.e. gap) between the lowest and the second band can be denoted by
$\Delta(\{t_{\nu},\epsilon_{\nu}\})$. Choosing zero energy scale, 
the presence of a bare flat band means 
$A(\{0,t_{\nu},\epsilon_{\nu}\})= B(\{0,t_{\nu},\epsilon_{\nu}\})=0$, the placement 
of the flat band in the lowest position representing a supplementary condition
$F(\{t_{\nu},\epsilon_{\nu}\})=0$. Furthermore, the presence of a contact point
between the lowest flat band and the dispersive band situated just above can be
simply given by $\Delta(\{t_{\nu},\epsilon_{\nu}\})=0$. These four equations
(i.e. $A=0,B=0,F=0,\Delta=0$) always provide solutions where the number of
Hamiltonian parameters in $\hat H_0$ -- as in realistic cases (e.g., in the 
studied poly(3-alkylthiophene) polymer pentagon chain case is seven) -- is high.

\section{Acknowledgments}

For M. Gul\'acsi this research was realized in the frames of 
TAMOP 4.2.4. A/2-11-1-2012-0001 
"National Excellence Program - Elaborating and operating an inland student 
and researcher personal support system". The project was subsidized by the 
European Union and co-financed by the European Social Fund. 
Z. Gul\'acsi kindly acknowledges financial support provided by the Alexander von 
Humboldt Foundation, OTKA-K-100288 (Hungarian Research Funds for Basic 
Research) and TAMOP 4.2.2/A-11/1/KONV-2012-0036 (co-financed by EU and European 
Social Fund). 
                                         

\bibliographystyle{model1a-num-names}
\bibliography{<your-bib-database>}

\begin{thebibliography}{00}
\bibitem{Explic1}
Note that for $N<N_c$ the ground state (\ref{E19}) becomes a sum of the 
form $\sum_{{\bf i}_1,...,{\bf i}_N} \alpha_{{\bf i}_1,...,{\bf i}_N} \prod_{{\bf i}=1}^{N} 
\hat B^{\dagger}_{{\bf i},\sigma_{\bf i}} |0\rangle$, where the product is made over
all possible $({\bf i}_1,...,{\bf i}_N)$ sets. Because the connectivity is 
missing, this state is paramagnetic.
\bibitem{Explic3}
In higher dimensions, the bare band structure can emerge in the form
$\sum_{n=0}^m A_n \cos^n {\bf k}{\bf a} =0$, where $m$ is finite. In this case, 
the presence of a flat band can be given as $A_n=0$ for all $n$.
\end{thebibliography}

\end{document}